\documentclass[10pt]{article}

\begin{document}

\title{Twisting type-N vacuum fields with a group $H_2$}
\author{
F.J. Chinea\footnote{E-mail address: chinea@eucmos.sim.ucm.es} \, and F. Navarro-L\'erida\footnote{E-mail address: fnavarro@eucmos.sim.ucm.es} \\ 
Dept. de F\'{\i}sica Te\'orica II,  Ciencias F\'{\i}sicas, \\
Universidad Complutense de Madrid \\
E-28040 Madrid, Spain}

\date{}
\maketitle

\begin{abstract}
We derive the equations corresponding to twisting type-N vacuum gravitational fields with one Killing vector and one homothetic Killing vector by using the same approach as that developed by one of us in order to treat the case with two non-commuting Killing vectors. We study the case when the homothetic para\-me\-ter $\phi$ takes the value $-1$, which is shown to admit a reduction to a third-order real ordinary differential equation for this problem, similar to that previously obtained by one of us when two Killing vectors are present. 
\end{abstract}

\vspace{0.5cm}

\hspace{0.2cm} PACS numbers: 0420C, 0430N

\section{Introduction}

The problem of finding exact solutions of the vacuum Einstein equations with algebraic type N and twisting rays appears to be a difficult one, as the only known field of such a type was given a quarter of a century ago by Hauser \cite{Hauser}, and no new ones have appeared in the meantime. Aside from its obvious physical interest as a model for asymptotic gravitational radiation, some remarkable mathematical structures  and results emerge from an analysis of the problem. Naturally, at the present stage it is appropriate to assume the existence of symmetries in order to simplify the equations. The case of a group $H_2$ (corresponding to the existence of two homothetic Killing vectors) is particularly suitable, especially in view of the fact that Hauser's solution is of such a type. Reductions of the equations in such a situation have been considered by a number of authors \cite{Collinson}, \cite{McIntosh1}, \cite{Herlt}, \cite{Ludwig1},
and \cite{Finley1}. The special case of a two-parameter iso\-me\-try group (i.e. when both homothetic Killing vectors are Killing vectors in the strict sense) has been treated in \cite{Javier3}  (where it was shown that, in that case, the equations can be reduced  to a final third-order real ordinary di\-ffe\-ren\-tial equation) and \cite{Javier4}, where a new first integral of the equations was found (no such additional first integral has yet been found in the general $H_2$ case). In this paper, we enlarge somewhat the scope of the previous papers \cite{Javier3} and \cite{Javier4}, in order to write down a similar reduction in the $H_2$ case, by using the techniques introduced in those references. In the parti\-cu\-lar case where the homothetic parameter has the value $\phi=-1$, we are able to perform a final reduction to a single third-order real differential equation, analogous to the reduction to a final third-order equation introduced in  \cite{Javier3} in the case $\phi=0$.

\section{Reduction of the equations in the presence of an \boldmath{$H_2$}}
Let $\{m, \bar{m}, l, n\}$ be a null tetrad of 1-forms such that the metric can be written as
\begin{equation}
g =2  m \otimes_s\bar{m} -2 l \otimes_s n, \label{eq_1}
\end{equation}
where $\otimes_s$ denotes the symmetrized tensor product and the bar denotes complex conjugation. As indicated in \cite{Javier1}, the Einstein equations in vacuum may be formulated in a very compact form by introducing the matrix-valued 1 forms
\begin{equation} \displaystyle
\eta = \left( \begin{array}{cc}
l&m\\
\bar{m}&n
\end{array} \right) , \quad 
\gamma = \left( \begin{array}{cc}
\frac{1}{2} \alpha&-\beta\\
\epsilon&-\frac{1}{2} \alpha
\end{array} \right). \label{eq_2}
\end{equation}  
By using them, the field equations take the following form:
\begin{equation}
d \eta = \gamma \wedge \eta - \eta \wedge \gamma^{\dag}, \label{eq_3}
\end{equation}
\begin{equation}
(d {\gamma} - \gamma \wedge \gamma) \wedge \eta = 0, \label{eq_4'}
\end{equation}
where the dagger denotes Hermitian conjugation. Equation (\ref{eq_4'}) is equivalent to
\begin{equation}
d \gamma - \gamma \wedge \gamma = W \wedge \eta \equiv R, \label{eq_4}
\end{equation}
with 
\begin{equation} \displaystyle
W = \left( \begin{array}{cc}
\Psi_2 (n-l)-\Psi_3 m + \Psi_1 \bar{m}&- \Psi_1 (n-l) + \Psi_2 m -\Psi_0 \bar{m}\\
\Psi_3 (n-l)- \Psi_4 m + \Psi_2 \bar{m}&-\Psi_2 (n-l) + \Psi_3 m - \Psi_1 \bar{m}  
\end{array} \right), \label{eq_5} \end{equation}
$\{ \Psi_0, \Psi_1, \Psi_2, \Psi_3, \Psi_4 \}$ being the components of the Weyl spinor. \par
\indent We are interested in the case when the metric (\ref{eq_1}) is of type N. The 1-form $l$ is the repeated principal null eigenform, and its twist is assumed not to vanish. By means of an appropriate gauge transformation, $\gamma$ can be brought to the form 
\begin{equation}
\gamma =  \left( \begin{array}{cc}
0&d \zeta\\
w d \zeta&0
\end{array} \right), \label{eq_6} \end{equation}
where $\zeta$ and $w$ are complex variables \cite{Javier2}.\par
\indent Now, let us consider the presence of symmetries in our metric. If $v$ is a homothetic Killing vector field such that $ {\cal L}_v g = 2 \phi_v g$, with $\phi_v$ a constant, it is very easy to check that the following relations hold:
\begin{eqnarray}
&&{\cal L}_v \eta = \chi_v \eta + \eta \chi_v^{\dag}, \label{eq_7}\\
&&{\cal L}_v \gamma = d \chi_v + [ \chi_v, \gamma ], \label{eq_8}\\
&&{\cal L}_v R = [\chi_v, R ], \label{eq_9}
\end{eqnarray} 
where
\begin{equation}
\chi_v =   \left( \begin{array}{cc}
a_v&b_v\\
c_v&-a_v+\phi_v\\
\end{array} \right), \label{eq_10} \end{equation}
with $a_v, b_v,$ and $c_v$ complex functions (the case of a Killing vector field corresponds to $\phi_v=0$, \cite{Javier2}). 

Let us suppose that the metric admits a homothetic Killing vector field $h$ ($ {\cal L}_h g = 2 \phi g$) and a Killing vector field $\xi$ such that they both constitute a symmetry group $H_2$, with the commutation relation
\begin{equation}
[\xi, h] = \lambda \xi + \mu h, \quad \lambda, \mu \mbox{ constants.} \label{eq_11}
\end{equation}
Following \cite{Javier2} and using the coordinate system $\{\sigma, u, \zeta, \bar{\zeta} \}$, where $\sigma$ and $u$ are real and $w= w(u,\zeta,\bar{\zeta})$ (it should be noticed that $\sigma$ was called $v$ in \cite{Javier2}), it can be shown that $\xi$ and $h$ are given by
\begin{eqnarray} \left\{ \begin{array}{c} \displaystyle
\xi = \xi^{\sigma} \partial_{\sigma} +\frac{1}{w_{,u}} \Bigl[ \frac{1}{2} \xi_{,\zeta \zeta \zeta}^{\zeta} - 2 w \xi_{, \zeta}^{\zeta} -w_{, \zeta} \xi^{\zeta} - w_{, \bar{\zeta}} \xi^{\bar{\zeta}} \Bigr] \, \partial_u + \xi^{\zeta} \partial_{\zeta} + \xi^{\bar{\zeta}} \partial_{\bar{\zeta}}, \label{eq_12}  \bigskip
\\
\chi_{\xi} =   \left( \begin{array}{cc} 
\frac{1}{2} \xi_{,\zeta}^{\zeta}&0\\  
\frac{1}{2} \xi_{, \zeta \zeta}^{\zeta}&-\frac{1}{2} \xi_{,\zeta}^{\zeta}
\end{array} \right), \label{eq_13}  \bigskip\\ 
\xi^{\zeta} = \xi^{\zeta}(\zeta), \quad \xi^{\bar{\zeta}} = \overline{\xi^{\zeta}}, \quad \xi^{\sigma} = \overline{\xi^{\sigma}}, 
\end{array} \right.
\end{eqnarray}
where $\xi^{\sigma}$ may depend on $\{\sigma, u, \zeta, {\bar \zeta}\}$, and  

\begin{eqnarray} \left\{ \begin{array}{c} \displaystyle
h = h^{\sigma} \partial_{\sigma} +\frac{1}{ w_{,u}} \Bigl[ \frac{1}{2} h_{,\zeta \zeta \zeta}^{\zeta} - 2 w h_{, \zeta}^{\zeta} -w_{, \zeta} h^{\zeta} - w_{, \bar{\zeta}} h^{\bar{\zeta}} \Bigr] \, \partial_u + h^{\zeta} \partial_{\zeta} + h^{\bar{\zeta}} \partial_{\bar{\zeta}}, \label{eq_14} \bigskip
\\
\chi_h =   \left( \begin{array}{cc} 
\frac{1}{2} ( h_{,\zeta}^{\zeta}+ \phi)&0\\ 
\frac{1}{2} h_{, \zeta \zeta}^{\zeta}&-\frac{1}{2} ( h_{,\zeta}^{\zeta} - \phi)
\end{array} \right), \label{eq_15} \bigskip \\ 
h^{\zeta} = h^{\zeta}(\zeta), \quad h^{\bar{\zeta}} = \overline{h^{\zeta}}, \quad h^{\sigma} = \overline{h^{\sigma}},
\end{array} \right.
\end{eqnarray}
where $h^{\sigma}$ may depend on $\{\sigma, u, \zeta, {\bar \zeta}\}$, too. These expressions are a consequence of equations (\ref{eq_8}) and (\ref{eq_9}), when $v$ is subs\-ti\-tu\-ted by $\xi$ and $h$, respectively. \par
\indent The relation (\ref{eq_11}) can be easily worked out to be 
\begin{equation}
{\cal L}_{\xi} \chi_{h} - {\cal L}_h \chi_{\xi} = [ \chi_{\xi}, \chi_h ] + \lambda \chi_{\xi} + \mu \chi_h. \label{eq_16}
\end{equation} 
When we impose the previous relation on $\xi$ and $h$, we obtain
\begin{equation}
\xi^{\zeta} h_{, \zeta \zeta}^{\zeta} - h^{\zeta} \xi_{, \zeta \zeta}^{\zeta} = \lambda \xi_{, \zeta}^{\zeta} + \mu (h_{, \zeta}^{\zeta} + \phi), \label{eq_17}
\end{equation}
\begin{equation}
\xi^{\zeta} h_{, \zeta \zeta}^{\zeta} - h^{\zeta} \xi_{, \zeta \zeta}^{\zeta} = \lambda \xi_{, \zeta}^{\zeta} + \mu (h_{, \zeta}^{\zeta} - \phi). \label{eq_18}
\end{equation}
If $\phi \neq 0$ (when $\phi=0$ it is easy to realize that we can suppose that $[ \xi, h ] = \xi$), by subtracting (\ref{eq_17}) from (\ref{eq_18}) we can conclude that $\mu=0$, giving the relation
\begin{equation}
\xi^{\zeta} h_{, \zeta \zeta}^{\zeta} - h^{\zeta} \xi_{, \zeta \zeta}^{\zeta} = \lambda \xi_{, \zeta}^{\zeta}. \label{eq_19}
\end{equation}
Redefining our symmetry vector fields and the homothetic constant as $\tilde{\xi} = \xi$, $\tilde{h} = \lambda^{-1}
h$, and $\tilde{\phi} = \lambda \phi$ and dropping tildes,
we can write
 
\begin{equation} \left\{ \begin{array}{l} \displaystyle
[\xi, h] = \xi, \label{eq_20}
\\
{\cal L}_{\xi} g = 0, \quad {\cal L}_h g = 2 \phi g, \label{eq_21} 
\\
\xi^{\zeta} h_{, \zeta \zeta}^{\zeta} - h^{\zeta} \xi_{, \zeta \zeta}^{\zeta} =  \xi_{, \zeta}^{\zeta}. \label{eq_22}
\end{array} \right.
\end{equation}

We have obtained the same relations as in the case of two non-commuting Killing vectors, so we can proceed similarly in order to get the following {\it cano\-ni\-cal} form of the symmetry fields:
\begin{equation}
\xi = i \partial_{\zeta} - i \partial_{\bar{\zeta}}, \label{eq_23}
\end{equation}
\begin{equation}
h = \zeta \partial_{\zeta} + \bar{\zeta} \partial_{\bar{\zeta}}, \label{eq_24}
\end{equation}
and 
\begin{equation}
w = F (u) (\zeta + \bar{\zeta})^{-2},
\end{equation}
where $F(u)$ is an arbitrary complex function \cite{Javier2}.\par
\indent Then, using now (\ref{eq_3}), (\ref{eq_4'}),
(\ref{eq_6}), and (\ref{eq_7}), when $v$ equals $\xi$ and
$h$, and 
\begin{equation} \displaystyle
\chi_{\xi} = \left( \begin{array}{cc}
0&0\\
0&0
\end{array} \right), \quad  \chi_h =\frac{1}{2} \left(\begin{array}{cc}
\phi +1&0\\
0&\phi -1
\end{array}\right), \label{eq_26}
\end{equation}
the tetrad 1-forms $\{m, \bar{m}, l, n\}$ can be written as
\begin{eqnarray} \displaystyle
l &=& (\zeta+\bar{\zeta})^{\phi+1} d u + D (\zeta+\bar{\zeta})^{\phi} d \zeta + \bar{D}  (\zeta+\bar{\zeta})^{\phi} d \bar{\zeta}, \label{eq_27} \\
n&=& (\zeta+\bar{\zeta})^{\phi-1} d \sigma + \frac{1}{2} \phi (M D^{-1} +\bar{M}{\bar{D}}^{-1}) (\zeta+\bar{\zeta})^{\phi-1} d u \nonumber\\
&&\mbox{} + \{ (\phi-1) [\sigma + M +\frac{1}{2} \phi (D-\bar{D})] + F\bar{D} \} (\zeta+\bar{\zeta})^{\phi-2} d \zeta \label{eq_28} \\
&&\mbox{} + \{ (\phi-1) [\sigma + \bar{M} -\frac{1}{2} \phi (D-\bar{D})] + \bar{F}D \} (\zeta+\bar{\zeta})^{\phi-2} d \bar{\zeta}, \nonumber\\
m &=& \bar{M} {\bar{D}}^{-1} (\zeta+\bar{\zeta})^{\phi} d u + [-\sigma + \frac{1}{2} \phi (D - \bar{D})] (\zeta+\bar{\zeta})^{\phi-1} d \zeta \nonumber\\
&&\mbox{} + \bar{M} (\zeta+\bar{\zeta})^{\phi-1} d \bar{\zeta}, \label{eq_29}
\end{eqnarray}
where $D$, $F$, and $M$ are complex functions of the
variable $u$ only, which verify the following system of
equations:
\begin{eqnarray}
&&M_u = -F + \phi M D^{-1}, \label{eq_30}\\
&&F_u = -2 F \bar{D}^{-1}, \label{eq_31}\\
&&D_u = (\phi+1) - M D^{-1}, \label{eq_32}\\
&&(\phi-2) [ \phi (\phi-1) (D - \bar{D}) + (\phi-1) (M-\bar{M}) \nonumber\\
&&\mbox{} -\bar{F} D + F \bar{D}] + F \bar{M} -\bar{F} M = 0, \label{eq_33}
\end{eqnarray} 
which clearly generalizes the case discussed in \cite{Javier3}, which is simply the case $\phi=0$.\par
\indent The only non-vanishing spin coefficients are
\begin{eqnarray} \displaystyle
\lambda &=& \frac{F \bar{D}}{[\sigma \bar{D}+ \bar{M}D -\frac{1}{2} \phi \bar{D} (D- \bar{D})] (\zeta + \bar{\zeta})^{\phi+1}},  \label{eq_34}\\
\nu &=& \frac{F \bar{M}}{[\sigma \bar{D}+ \bar{M}D -\frac{1}{2} \phi \bar{D} (D- \bar{D})] (\zeta + \bar{\zeta})^{\phi+2}}, \label{eq_35}\\
\rho &=& - \frac{\bar{D}}{[\sigma \bar{D}+ \bar{M}D -\frac{1}{2} \phi \bar{D} (D- \bar{D})] (\zeta + \bar{\zeta})^{\phi-1}}, \label{eq_36}\\
\tau &=&- \frac{M}{[\sigma \bar{D}+ \bar{M}D -\frac{1}{2} \phi \bar{D} (D- \bar{D})] (\zeta + \bar{\zeta})^{\phi}}.  \label{eq_37}
\end{eqnarray}

The component $\Psi_4$ of the Weyl spinor (the other ones vanish) is given by
\begin{equation} \displaystyle
\Psi_4 = \frac{2 F}{[\sigma \bar{D}+ \bar{M}D -\frac{1}{2} \phi \bar{D} (D- \bar{D})] (\zeta + \bar{\zeta})^{2(\phi+1)}}.  \label{eq_38}
\end{equation}
Note that we must have $F \neq 0$ if we want the spacetime not to be flat.\par
\indent As for the non-vanishing twist condition, $\rho \neq \bar{\rho}$, it can be expressed as
\begin{equation}
M {\bar{D}}^2 - \bar{M} D^2 + \phi D \bar{D} (D- \bar{D}) \neq 0. \label{eq_39}
\end{equation}
\indent In the special case $\phi=0$, all these equations trivially reduce to those given in \cite{Javier4}. \par

\section{The case with two Killing vectors
(\boldmath{$\phi=0$})}

It was shown \cite{Javier3} that the system of equations (\ref{eq_30})-(\ref{eq_33}) can be reduced to a single third-order real differential equation when $\phi=0$. The procedure was the following. Firstly, a new independent variable $v$ is defined as
\begin{equation}
v(u) \equiv \int F(u) {\bar F} (u) du. \label{J_eq_1}
\end{equation}
By using it instead of $u$, the system (\ref{eq_30})-(\ref{eq_33}) (with $\phi=0$) is transformed into 
\newpage
\begin{eqnarray} \displaystyle
&&M' = -{\bar F}^{-1}, \label{J_eq_2}\\
&&F' = -2 {\bar F}^{-1} {\bar D}^{-1}, \label{J_eq_3}\\
&&D' = (1 - M D^{-1}) F^{-1} {\bar F}^{-1}, \label{J_eq_4}\\
&&v = 2 F {\bar D} - 2M - F {\bar M} \label{J_eq_5}
\end{eqnarray}
(primes will denote derivatives with respect to $v$). Solving the previous system is equivalent to finding a solution of the second-order differential equation
\begin{equation} \displaystyle
\omega''= \frac{2 {\omega'}^2 {\bar \omega}'}{\omega - v \omega' - {\bar \omega} \omega'}, \label{J_eq_6}
\end{equation}
where $\omega = 2 M$; $D$ and $F$ can be obtained from (\ref{J_eq_2})-(\ref{J_eq_5}) in an algebraic form once (\ref{J_eq_6}) is solved. Following the steps indicated in  \cite{Javier3}, equation (\ref{J_eq_6}) can be ``linearized'' by means of a Legendre transformation, namely, $h \equiv f -v f'$, $x \equiv f'$, and $v = - h_x$, with $\omega = 2 f + 2 i g$, and $\omega' = 2x + 2iy$ ($f$, $g$, $x$, and $y$ real). Equation (\ref{J_eq_6}) is then equivalent to the set of equations:
\begin{eqnarray}
&& g_x + y h_{xx} =0, \label{J_eq_7}\\
&& A g + B h + L h_x + Q h_{xx} = 0, \label{J_eq_8}\\
&& a g + b h + l h_x + q h_{xx} = 0, \label{J_eq_9}
\end{eqnarray} 
where
\begin{eqnarray}  \left\{ \begin{array}{ll} \displaystyle
A=\frac{y}{4 (x^2+y^2)^2}, \quad & \displaystyle  a= \frac{x+2 (x^2+y^2)}{4 (x^2+y^2)^2}, \label{J_eq_10}\\
\displaystyle B= \frac{x-2 (x^2+y^2)}{4 (x^2+y^2)^2}, \quad & \displaystyle b=\frac{-y}{4 (x^2+y^2)^2}, \label{J_eq_11}\\
 \displaystyle L = \frac{y^2+2 x (x^2+y^2)}{4 (x^2+y^2)^2}, \quad & \displaystyle l=\frac{x y}{4 (x^2+y^2)^2}, \label{J_eq_12}\\
 \displaystyle Q=\frac{1}{1+y_x^2}, \quad & \displaystyle q=\frac{-y_x}{1+y_x^2}. \label{J_eq_13}
\end{array} \right.
\end{eqnarray}  
Assuming $Lq-lQ \neq 0$, we can solve (\ref{J_eq_7})-(\ref{J_eq_9}) for $h_x$ and $h_{xx}$ in the form
\begin{eqnarray}
&&h_x = S g + T h \label{J_eq_14},\\
&&h_{xx} = s g + t h \label{J_eq_15},
\end{eqnarray}
where $S$, $T$, $s$, and $t$ are functions of $x$, $y$, and
$y_x$. If we differentiate (\ref{J_eq_14}) with respect to
$x$ and set the result equal to (\ref{J_eq_15}), we obtain
\begin{equation} \displaystyle
h = \frac{S_x - y s S + T S -s}{t +y t S - T^2- T_x} g \equiv V g. \label{J_eq_16}
\end{equation}
Now, we differentiate this equation with respect to $x$. Then we use (\ref{J_eq_7}) in order to eliminate $g_x$ from the result, (\ref{J_eq_14}) and (\ref{J_eq_15}) to eliminate $h_x$ and $h_{xx}$, and (\ref{J_eq_16}) to substitute $g$ as a function of $h$. As $h$ cannot be identically zero, we obtain the following ordinary differential equation of the third order for the real function $y(x)$:
\begin{equation}
V_x - y s V -y t V^2- S - T V =0. \label{J_eq_17}
\end{equation}\par
\indent The resulting third-order equation was not explicitly written down in \cite{Javier3} due to its length: at that time, with symbolic manipulation systems having a limited factoring ability, the full equation expanded in monomials had more than 200 terms! However, it can be shown that the equation factors into a short factor and a larger one. Unfortunately, it is easy to show that the short factor cannot vanish in the non-trivial case (i.e. non-flat space, with a non-vanishing twist), so we are left with the (still complicated, but printable) third-order equation
 \begin{equation} \begin{array}{l}
4 r^4 \{4 r^2 y_x+\delta [y (1-y_x^2)-2 x y_x]\} y_{xxx}+12 r^4 [\delta (x+y y_x)-2 r^2] y_{xx}^2\\
\mbox{}-4 r^2 \{5 (y^2 y_x^3+3 x y y_x^2+3 x^2 y_x-x y) \delta-[(7+8 x) y_x^3+12 y y_x^2\\
\mbox{}+12 \delta y_x-8 y] r^2\} y_{xx}+(1+y_x^2) \{[(1-7 x+6 x^2) r^2-x^2 \delta^2] y_x^4\\
\mbox{}+y [(7-12 x) r^2+4 x \delta^2] y_x^3+[6 r^4-6 y^2 \delta^2+(4+x) r^2] y_x^2\\
\mbox{}+y [(23+52 x) r^2-4 x \delta^2] y_x+y^2 (\delta^2-10 r^2)\} =0,

\end{array} \label{J_eq_18}
\end{equation}
where $r \equiv \sqrt{x^2+y^2}$ and $\delta \equiv 1+4 x$.
\par
\indent Once (\ref{J_eq_18}) is solved, we can substitute $y=y(x)$ into (\ref{J_eq_16}) to solve for $g$ and introduce this $g$ into (\ref{J_eq_14}) from which we obtain $h$ (and hence $g$) by a quadrature.\par

\indent It should be noted that several other third-order equations have subsequently appeared in the literature related to this subject. Similar techniques have been used in \cite{Herlt} and \cite{Ludwig1} in the case of the existence of one Killing vector and one homothetic Killing vector, also showing a reduction to a final, third-order equation. It is remarkable that the simplest such equation was found in \cite{Herlt} for the case $\phi=1$. The reduction to a third-order equation and the relation among different approaches have been further investigated in \cite{Finley1}, \cite{Finley2}, \cite{Ludwig2}, and \cite{Ludwig3} (see also references quoted therein and in \cite{Javier4}). Reference \cite{McIntosh2} gives a very useful review on this subject. 
\section{The case \boldmath{$\phi=-1$}}

Now we consider the case when $\phi=-1$ and we will see that it is possible to give a treatment similar to that used in \cite{Javier3}. For general $\phi$, let us introduce a new real independent variable $v$ (which generalizes the one introduced in the previous section for the special case $\phi =0$), defined by
\begin{equation}
v(u) \equiv \int [ F(u) \bar{F}(u) - \phi (\phi^2-1)(\phi -2) ] \, d u. \label{eq_40}
\end{equation}
In terms of the new independent variable $v$, the system (\ref{eq_30})-(\ref{eq_33}) can be written as
\begin{eqnarray} \displaystyle
&&M' = \frac{-F + \phi M D^{-1}}{F \bar{F} -\phi (\phi^2-1)(\phi-2)}, \label{eq_41}\\ 
&&F' = \frac{- 2 F {\bar{D}}^{-1}}{F \bar{F} -\phi (\phi^2-1)(\phi-2)}, \label{eq_42}\\
&&D' = \frac{(\phi+1)-MD^{-1}}{F \bar{F} -\phi (\phi^2-1)(\phi-2)}, \label{eq_43}\\
&&v = -(\phi-2) [ (\phi-1) (M+\phi D) + F \bar{D}] -F\bar{M} \label{eq_44}
\end{eqnarray}
(primes will denote derivatives with respect to $v$ again). It is convenient to define a new dependent variable $N$ instead of $M$:
\begin{equation}
N \equiv M + \phi D. \label{eq_44'}
\end{equation}
Equations (\ref{eq_41})-(\ref{eq_44}) take the following form:
\begin{eqnarray} \displaystyle
&&N' =  \frac{-F + \phi (\phi+1)}{F \bar{F} -\phi (\phi^2-1)(\phi-2)}, \label{eq_45}\\ 
&&F' = \frac{- 2 F {\bar{D}}^{-1}}{F \bar{F} -\phi (\phi^2-1)(\phi-2)}, \label{eq_46}\\
&&D' = \frac{(2 \phi+1)-N D^{-1}}{F \bar{F} -\phi (\phi^2-1)(\phi-2)}, \label{eq_47}\\
&&v =2 F \bar{D} - (\phi-2)(\phi-1) N - F \bar{N}. \label{eq_48}
\end{eqnarray}
%
\indent As may be seen in expression (\ref{eq_40}), there are three obvious special cases (aside from $\phi=0$) in that transformation, namely, $\phi=1, 2, -1.$ The first one corresponds to Herlt's case
$N=2$ \cite{Herlt}, where $N=\phi+1$; as for the second one,
we are not aware of any previous explicit reference to it in
the literature. For these two cases we have not been able to
find a reduction similar to that described in  the
previous section, so we are going to concentrate on the
value $-1$.\par  
%
%
\indent If we set $\phi$ to be equal to $-1$, then
\begin{eqnarray}
&&N' = - {\bar{F}}^{-1}, \label{eq_49}\\
&&F' = -2 {\bar{F}}^{-1} {\bar{D}}^{-1}, \label{eq_50}\\
&&D' = (-1-ND^{-1}) F^{-1} {\bar{F}}^{-1}, \label{eq_51}\\
&&v = 2 F \bar{D} -6 N -F \bar{N}. \label{eq_52}
\end{eqnarray}

It is easy to check that these equations can be used to explicitly solve for $D$ and $F$ as functions of $N$, $\bar{N}$,
$N'$, and $\bar{N}'$; the remaining function $N$ is forced
to satisfy a single, second-order ordinary differential
equation. Introducing $\Omega \equiv 2 N$ for convenience,
the resulting equation is
\begin{equation} \displaystyle
\Omega'' =  \frac{ 2 {\Omega'}^2 {\bar{\Omega}}'}{\Omega - v \Omega' - 3 {\bar \Omega} \Omega'}. \label{eq_53}
\end{equation}
Note that it is almost identical to equation (\ref{J_eq_6}), except for the factor of three in the denominator. For this reason, we can proceed similarly to the case with two Killing vectors and make a Legendre transformation, according to $\Omega \equiv 2 f + 2 i g$, $\Omega' \equiv  2x + 2 i y$ ($f$, $g$, $x$, and $y$ real), with $h \equiv f - v f'$. Proceeding in this way, we obtain the following set of equations, linear in $g$ and $h$:
\begin{eqnarray}
&& g_x + y h_{xx} =0, \label{eq_54}\\
&& A g + B h + L h_x + Q h_{xx} = 0, \label{eq_55}\\
&& a g + b h + l h_x + q h_{xx} = 0, \label{eq_56}
\end{eqnarray} 
where
\begin{eqnarray}  \left\{ \begin{array}{ll} \displaystyle
A=\frac{y}{4 (x^2+y^2)^2}, \quad & \displaystyle  a= \frac{x+6 (x^2+y^2)}{4 (x^2+y^2)^2}, \label{eq_57}\\
\displaystyle B= \frac{x-6 (x^2+y^2)}{4 (x^2+y^2)^2}, \quad & \displaystyle b=\frac{-y}{4 (x^2+y^2)^2}, \label{eq_58}\\
 \displaystyle L = \frac{y^2+6 x (x^2+y^2)}{4 (x^2+y^2)^2}, \quad & \displaystyle l=\frac{x y}{4 (x^2+y^2)^2}, \label{eq_59}\\
 \displaystyle Q=\frac{1}{1+y_x^2}, \quad & \displaystyle q=\frac{-y_x}{1+y_x^2}. \label{eq_60}
\end{array} \right.
\end{eqnarray}
As can be seen, everything is formally similar to the case $\phi=0$. Then, by applying the compatibility conditions for the system (\ref{eq_54})-(\ref{eq_56}), and proceeding as in the previous section, it is found that the problem for the case $\phi=-1$ reduces to the following real ordinary differential equation of third order for the real function $y(x)$

\begin{equation} \begin{array}{l}

4 r^4 \{12 r^2 y_x+\Delta [y (1-y_x^2)-2 x y_x]\} y_{xxx}+12 r^4 [\Delta (x+y y_x)-6 r^2] y_{xx}^2\\
\mbox{}-4 r^2 \{7 (y^2 y_x^3+3 x y y_x^2+3 x^2 y_x-x y) \Delta-[(11+48 x) y_x^3+36 y y_x^2\\
\mbox{}+18 \Delta y_x-48 y] r^2\} y_{xx}+(1+y_x^2) \{[(1-11 x+18 x^2) r^2-x^2 \Delta^2] y_x^4\\
\mbox{}+y [(11-36 x) r^2+4 x \Delta^2] y_x^3+[18 r^4-6 y^2 \Delta^2+(4+13 x) r^2] y_x^2\\
\mbox{}+y [(59+540 x) r^2-4 x \Delta^2] y_x+y^2 (\Delta^2-126 r^2)\}=0,

\end{array} \label{eq_65}
\end{equation}
where $r \equiv \sqrt{x^2+y^2}$ and $\Delta \equiv 1+12 x$.
Once (\ref{eq_65}) is solved, $h$ is given by a quadrature,
and $g$ is given in finite terms. 
\section{Hauser's solution}

Finally, for the sake of completeness, we write down the explicit expression of Hauser's solution \cite{Hauser} by using this approach. The solution corresponds to the case $\phi=5/2$ and assumes the following ansatz for the function $F$:

\begin{equation}
F= \frac{3}{2 (1+i z)}, \label{eq_66}
\end{equation}
where $z$ is a real function of $u$. The functions $M$ and
$D$ may be obtained from (\ref{eq_31}) and (\ref{eq_32}) by
\begin{equation} \begin{array}{l}
D= \displaystyle  -2 {\bar{F}}{\bar{F}_{,u}}^{-1},\\
M=D (\phi +1 -D_{,u}), \end{array} \label{eq_67} 
\end{equation}  
with $\phi=5/2$. We only have to check equations
(\ref{eq_30}) and (\ref{eq_33}). However, equation (\ref{eq_33})
becomes an identity when we substitute $\phi=5/2$,
equation (\ref{eq_66}), and (\ref{eq_67}) into it. Then, if we impose
(\ref{eq_30}) on these functions $F$, $D$, and $M$, we can
see that it is verified if the function $z$ satisfies the
following differential equation:
\begin{equation} 
\displaystyle z_{uuu}= \displaystyle  \frac{3 {z_u}^3}{16 (1
+ z^2)} + 3 \frac{{z_{uu}}^2}{z_u}.
\label{eq_68}\end{equation} And finally, the standard
Hauser equation is obtained by defining $p \equiv
1/{z_u},$ and using $z$ as an independent variable:
\begin{equation}
\displaystyle  \frac{d^2 p}{dz^2} + \frac{3}{16 (1 +z^2)} p
=0. \label{eq_69}
\end{equation}

\section{Conclusions}

By using techniques similar to those introduced in the two-Killing-vector case \cite{Javier3} and \cite{Javier4}, we have given a reduction of the equations for twisting type-N vacuum fields with $H_2$ symmetry, in the gauge formerly used in the 2-Killing case. In the present more general context, we have given an explicit reduction to a final third-order, real, ordinary differential equation in the case with homothetic parameter $\phi=-1$, in close parallel with the third-order equation previously found when $\phi=0$.  

\section*{Acknowledgments}

We thank L. Fern\'andez-Jambrina, L. M. Gonz\'alez-Romero, and M. J. Pareja
for discussions. Financial support from Direcci\'on General de Ense\~nanza
Superior e Investigaci\'on Cient\'{\i}fica (project PB98-0772) is gratefully
acknowledged.


\end{document}